\documentclass[11pt]{article}
\usepackage{amssymb}
\usepackage{amsmath}
\usepackage{amscd}
\usepackage{latexsym}

\catcode`@=11
\renewcommand{\section}{\@startsection{section}{1}{0pt}{\medskipamount}
{\medskipamount}{\large\bf}}
\numberwithin{equation}{section}
\catcode`@=12

\def\b{\beta}

\def\la{\lambda}
\def\m{\mu}

\def\1{\bar 1}
\def\2{\bar 2}
\def\3{\bar 3}

\newcommand{\yb}{\bar{y}}

\newcommand{\zb}{\bar{z}}

\newcommand{\C}{\mathbb C}
\newcommand{\R}{\mathbb R}

\newcommand{\Zcal}{{\cal Z}}
\newcommand{\Acal}{{\cal A}}
\newcommand{\Ncal}{{\cal N}}
\newcommand{\Lcal}{{\cal L}}
\newcommand{\Fcal}{{\cal F}}
\newcommand{\Ecal}{{\cal E}}

\newcommand{\Ucal}{{\cal U}}

\def\im{\mbox{i}}
\def\N2{$N{=}2$}
\def\pa{\mbox{$\partial$}}
\def\diff{\mbox{d}}
\def\tr{{\rm tr}}
\def\sfrac#1#2{{\textstyle\frac{#1}{#2}}}
\def\>{\rangle}
\def\<{\langle}
\def\+{\dagger}
\def\={\ =\ }

\begin{document}

\begin{titlepage}
\setcounter{page}{0}
\begin{flushright}
ITP-UH-01/08
\end{flushright}

\vskip 3.0cm

\begin{center}

{\Large\bf 
Non-Abelian Vortices on Riemann Surfaces: an Integrable 
Case$^*$
   }

\vspace{15mm}
{\Large Alexander D. Popov}
\\[5mm]
\noindent {\em Institut f\"ur Theoretische Physik,
Leibniz Universit\"at Hannover \\
Appelstra\ss{}e 2, 30167 Hannover, Germany }
\\
{Email: {\tt popov@itp.uni-hannover.de}}
\\[2mm]
and
\\[2mm]
\noindent {\em Bogoliubov Laboratory of Theoretical Physics, JINR\\
141980 Dubna, Moscow Region, Russia}
\\
{Email: {\tt popov@theor.jinr.ru}}
\vspace{15mm}

\begin{abstract}
\noindent We consider U$(n+1)$ Yang-Mills instantons on the space $\Sigma\times S^2$,
where $\Sigma$ is a compact Riemann surface of genus $g$. Using an SU(2)-equivariant
dimensional reduction, we show that the U$(n+1)$ instanton equations on $\Sigma\times S^2$
are equivalent to non-Abelian vortex equations on $\Sigma$. Solutions to these  equations 
are given by  pairs $(A,\phi)$, where $A$ is a gauge potential of the group U($n$) and $\phi$
is a Higgs field in the fundamental representation of the group U($n$). We briefly compare this 
model with other non-Abelian Higgs models considered recently. Afterwards we show that 
for $g>1$, when $\Sigma\times S^2$ becomes a gravitational instanton, the non-Abelian vortex 
equations are the compatibility conditions of two linear equations (Lax pair) and therefore 
the standard methods of integrable systems can be applied for constructing their solutions. 
\end{abstract}
\end{center}

\vfill

\textwidth 6.5truein
\hrule width 5.cm

{\small
\noindent ${}^*$
Supported in part by the Deutsche Forschungsgemeinschaft.}

\end{titlepage}

\section{Introduction and summary}

\noindent The Abelian Higgs model on $\R\times\R^2$ in the Bogomolny regime admits 
static vortex solutions~\cite{Abrikosov:1956sx} which describe magnetic flux tubes 
(vortex strings) penetrating a two-dimensional superconductor. It is widely believed 
that (electric) vortex strings play an important role in the confinement of quarks. 
Many results known for the Abelian Higgs model were generalized to compact Riemann 
surfaces and to the non-Abelian case (see e.g.~\cite{Bradlow}-\cite{Popov:2007ms} 
and references therein). These generalizations are of interest for many reasons. 
One reason for having vortices on a compact Riemann surface $\Sigma$ is to control 
the thermodynamic limit $N\to\infty$ with the density of vortices 
$N/\{$area of $\Sigma\}$ 
fixed~\cite{Manton, Manton:1998kq}. In this case one can reduce the evaluation of the 
partition function of the gas of vortices to the computation of the volume of the moduli 
space~\cite{Manton, Manton:1998kq}. Note that the moduli space of vortices on Riemann 
surfaces is compact (contrary to the case $\Sigma=\R^2$) and it naturally appears in 
the description of Gromov-Witten and other topological 
invariants (see e.g.~\cite{Salamon, Taubes}). Also, vortices on $\Sigma$ and corresponding 
Yang-Mills instantons on $\Sigma\times S^2$ appear~\cite{Popov:2007ms} in recently proposed 
twistor string theory~\cite{Witten, Berkovits}. On the other hand, non-Abelian vortices 
are actively studied presently since the confinement of quarks can be related with a 
condensation of non-Abelian monopoles and a non-Abelian analog of dual Meissner effect 
(see e.g.~\cite{group, group01, group1}). 

In this note, we consider U($n$) Yang-Mills-Higgs theory on a Riemann surface $\Sigma$ with 
a Higgs field in the fundamental representation of the group U($n$). Contrary to the flat 
space $\R^2$, where such theory reduces to the Abelian one, on Riemann surfaces even the 
simplest model with one Higgs field (one flavour) is really non-Abelian. 
Experience with other models shows that it is useful to know solutions exactly
- this can facilitate the analysis of their properties, their moduli space etc. In fact,
there exist well-developed solution-generating techniques for so-called integrable models. 
We show that the vortex equations on $\Sigma$ are integrable under certain conditions.
Namely, it is shown that non-Abelian vortices on $\Sigma$ can be identified with SO(3)-symmetric
Yang-Mills instantons on $\Sigma\times S^2$. Furthermore, for the genus $g>1$ one can choose a 
metric on  $M=\Sigma\times S^2$ such that the scalar curvature of $M$ vanishes and this manifold 
becomes~\cite{B} a (conformal) gravitational instanton~\cite{Penrose:1976js, Atiyah:1978wi}.
Then the Yang-Mills instanton equations on $\Sigma\times S^2$ become integrable as well as
the non-Abelian vortex equations on $\Sigma$. We describe geometrical and topological conditions 
for their integrability and write down the explicit Lax pair for the vortex equations.

\vspace{5mm}

\section{Manifolds $\Sigma\times\C P^1$}

Here we consider K\"ahler metrics on the product $\Sigma\times\C P^1$
of a real two-dimensional manifold  $\Sigma$ and the Riemann sphere $\C P^1$. 
As $\Sigma$ we will take the Euclidean space $\R^2$ or a compact Riemann surface
of genus $g\ge 0$. The case of $\R^2$ is kept for the Abelian subcase.

\noindent
{\bf Riemann surfaces.} For a Riemann surface $\Sigma$, the metric and the volume 
form are given in local (conformal) coordinates $z, \zb$ by
\begin{equation}\label{2.1}
\diff s^2_{\Sigma} = 2\,g_{z\zb}\,\diff z\,\diff \zb\quad\mbox{and}\quad
\omega_{\Sigma}^{} = \im\,g_{z\zb}\,\diff z\wedge\diff \zb\ ,
\end{equation}
respectively. Furthermore, for the nonvanishing components of the Christoffel symbol
and the Ricci tensor we have
\begin{equation}\label{2.2}
\Gamma^z_{zz} = 2\,\pa_z\log\rho\quad\mbox{and}\quad
\Gamma^{\zb}_{\zb\zb} = 2\,\pa_{\zb}\log\rho\quad\mbox{with}\quad
\rho^2:=g_{z\zb}\ ,
\end{equation}
\begin{equation}\label{2.3}
R_{z\zb} = -2\,\pa_z\pa_{\zb}\log\rho =\varkappa\, g_{z\zb}\quad\Longrightarrow\quad
R^{}_{\Sigma} = 2 g^{z\zb}R_{z\zb}=2\varkappa\ ,
\end{equation}
where $R^{}_{\Sigma}$ is the (constant) scalar curvature of $\Sigma$. In real local
coordinates $x^1, x^2$ on $\Sigma$ we have $z=x^1+\im x^2$ and $\zb=x^1-\im x^2$.
For $\R^2$ one should simply put $g_{z\zb}=\delta_{z\zb}$, $R_{z\zb}=0$ and $\varkappa =0$.

\smallskip

\noindent
{\bf Monopole bundle over $\C P^1$.} Formulae (\ref{2.1})-(\ref{2.3}) for the standard two-sphere 
$S^2\cong\C P^1$ of constant radius $R$ read
\begin{equation}\label{2.4}
\diff s_{\C P^1}^2 = \frac{4R^4}{(R^2+y\yb)^2}\,\diff y\,\diff \yb\quad\mbox{and}\quad
\omega_ {\C P^1}= - \frac{2\im\,R^4}{(R^2+y\yb)^2}\,\diff y\wedge\diff \yb\ ,
\end{equation}
\begin{equation}\label{2.5}
R_{y\yb}=-\pa_y \pa_{\yb}\log\tilde\rho^2 = \tilde\varkappa\, g_{y\yb}=\frac{1}{R^2}\,\tilde\rho^2
\quad\Rightarrow\quad R_{\C P^1}=\frac{2}{R^2}\quad\mbox{and}\quad\tilde\varkappa =\frac{1}{R^2}\ .
\end{equation}
Note that (\ref{2.4}) corresponds to the choice of orientation on $S^2$ inverse to the canonical one,
i.e. $y=x^3-\im x^4$ and $\yb=x^3+\im x^4$ for real local coordinates $x^3$, $x^4$
on $S^2$. 

Consider now the Hermitian complex line bundle\footnote{For more detailed description with
transition function etc. see e.g.~\cite {Popov:2004rt} and references therein.} $\Lcal\to\C P^1$
with a unitary Abelian connection $a$ and the curvature $f$ having the form
\begin{equation}\label{2.6}
a= \frac{1}{2(R^2+y\yb)}(\yb\,\diff y - y\,\diff\yb )\quad\mbox{and}\quad
f=\diff a =-\frac{R^2}{(R^2+y\yb)^2}\, \diff y\wedge\diff\yb = \frac{1}{2\im R^2}\,\omega_{\C P^1}\ .
\end{equation}
This is the Dirac monopole bundle over $\C P^1$.

\smallskip

\noindent
{\bf Gravitational instantons.} For the scalar curvature $R_M$ of the manifold $M=\Sigma\times\C P^1$
we have $R_M=2(\varkappa + \tilde\varkappa )$. Note that for
\begin{equation}\label{2.8}
\varkappa = - \tilde\varkappa = - \frac{1}{R^2}
\end{equation}
the K\"ahler manifold $M$ has self-dual Weyl curvature tensor~\cite{B}. Such manifolds are considered
as gravitational instantons in conformal gravity (see e.g.~\cite{Penrose:1976js, Atiyah:1978wi, Gibbons1}). 
At the same time, the manifold $M=\C P^1\times\C P^1$ with $\varkappa = \tilde\varkappa$ is a smooth 
Einstein manifold which can be considered as a gravitational instanton in Euclidean quantum gravity 
(see e.g.~\cite{Gibbons2} and references therein).

\vspace{5mm}

\section{Non-Abelian vortices on $\Sigma$ as Yang-Mills instantons on $\Sigma\times\C P^1$}

\noindent
{\bf SU(2)-equivariant gauge potential.} Consider the manifold $M=\Sigma\times\C P^1$.
Let $\Ecal\to M$ be an SU(2)-equivariant\footnote{This means a generalized SU(2)-invariance, 
i.e. invariance under space-time transformations up to gauge transformations~\cite{Forgacs:1979zs, 
Taubes:1979ps, GarciaPrada:1993qv}.} complex vector bundle of rank $r\ge2$ over $M$ with the 
group SU(2) acting trivially on $\Sigma$ and in the  standard way by SU(2)-isometry on 
$\C P^1=\ $SU(2)/U(1). Let $\Acal$ be a $u(r)$-valued local form of SU(2)-equivariant 
connection on $\Ecal$ (see e.g.~\cite{GarciaPrada:1993qv, Popov:2005ik}); it can be chosen 
in the form
\begin{equation}\label{3.1}
\Acal = \begin{pmatrix}A^1+ 1_{n}\cdot a& \sfrac{1}{\sqrt{2}}\phi\,\bar\b
\\-\sfrac{1}{\sqrt{2}}\phi^\+\,\b& A^2 - a\end{pmatrix}
\end{equation}
where $A^1$ and  $A^2$ are $u(n)$- and $u(1)$-valued gauge potentials on the 
rank $n$ and rank one vector bundles $E_1$ and $E_2$ over $\Sigma$ with $n+1=r$,
$\phi\in\,$Hom$(E_2, E_1)$ is a scalar transforming in bi-fundamental representation
$(n, \bar 1)$ of the group $U(n)\times U(1)$ and $\phi^\+$ is its Hermitian
conjugate. In (\ref{3.1}), $a$ is the one-monopole gauge potential on $\C P^1$ given in 
(\ref{2.6}) and
\begin{equation}\label{3.2}
\b:=\frac{\sqrt{2}\,R^2\,\diff y}{R^2+y\yb}\quad\mbox{and}\quad
\bar\b:=\frac{\sqrt{2}\,R^2\,\diff \yb}{R^2+y\yb}
\end{equation}
are forms on $\C P^1$ of type (1,0) and (0,1) satisfying
\begin{equation}\label{3.3}
\diff\bar\b + 2a\wedge\bar\b =0=\diff\b - 2a\wedge\b\quad\mbox{and}\quad 
\b\wedge\bar\b =\im\,\omega_{\C P^1}\ .
\end{equation}
Fields $A^1$, $A^2$ and $\phi$ depend only on coordinates of $\Sigma$  and the Higgs 
field $\phi$ can be identified with a section ($n\times 1$ column) of the bundle
$E:=E_1\otimes\bar E_2$.

\smallskip

\noindent
{\bf Symmetric field strength tensor.} In local complex coordinates on $\Sigma\times\C P^1$ 
the curvature $\Fcal =\diff\Acal + \Acal\wedge\Acal$ for $\Acal$ of the form (\ref{3.1}) has
the following field strength components:
\begin{subequations}\label{3.4}
\begin{eqnarray} 
\Fcal_{z\zb}&=&\begin{pmatrix} F_{z\zb}^1&0\\0&F_{z\zb}^2\end{pmatrix}\ ,\quad
\Fcal_{y\yb}=\begin{pmatrix}\frac{1}{2}g_{y\yb} (\frac{1}{R^2}\cdot 1_{n}-\phi\,\phi^{\+})&0\\
0&-\frac{1}{2}g_{y\yb} (\frac{1}{R^2} -\phi^{\+}\phi  )\end{pmatrix}\ ,\\
\Fcal_{\zb\yb}&=&\begin{pmatrix}0&\frac{\tilde\rho}{\sqrt{2}}\,(\pa_{\zb}\phi + 
A^1_{\zb}\phi - A^2_{\zb}\, \phi )\\0&0\end{pmatrix} = -(\Fcal_{zy})^{\+}\ ,
\end{eqnarray}
\end{subequations}
\begin{equation}\label{3.5} 
\Fcal_{z\yb}=\begin{pmatrix}0&\frac{\tilde\rho}{\sqrt{2}}\,(\pa_{z}\phi + A^1_{z}\phi -A^2_z\,\phi )\\
0&0\end{pmatrix} = -(\Fcal_{\zb y})^{\+}\ .
\end{equation}
Here  $\tilde\rho^2=g_{y\yb}$, $F^1=F^1_{z\zb}\,\diff z\wedge\diff\zb = \diff A^1 + A^1\wedge A^1$ and 
$F^2=F^2_{z\zb}\,\diff z\wedge\diff\zb = \diff A^2 $.

\smallskip

\noindent
{\bf Non-Abelian vortex equations on $\Sigma$.} Let us consider the Yang-Mills instanton
equations $\ast\Fcal =\Fcal$ on $M$, where $\ast$ is the Hodge operator. In local
coordinates on $\Sigma\times\C P^1$ these equations have the form
\begin{equation}\label{3.6}
\Fcal_{\zb\yb}=0=(\Fcal_{zy})^\+\quad\mbox{and}\quad 
g^{z\zb}\Fcal_{z\zb} + g^{y\yb}\Fcal_{y\yb}=0\ .
\end{equation}
Substitution of (\ref{3.4}) into (\ref{3.6}) shows that the instanton equations (\ref{3.6})
on $\Sigma\times\C P^1$ are equivalent to non-Abelian BPS\footnote{Namely, the first order 
Bogomolny equations~\cite{Bog} at critical value of the coupling constant.} vortex equations 
(cf.~\cite{Bradlow, GarciaPrada:1993qv, Popov:2005ik}) on $\Sigma$:
\begin{subequations}\label{3.7}
\begin{eqnarray}
&\pa_{\zb}\phi + (A^1_{\zb} - A^2_{\zb}\cdot 1_n )\phi =0\ ,\\
&2F^1_{z\zb}=g_{z\zb}\,\left(\frac{1}{R^2}\cdot 1_{n}-\phi\,\phi^\+ \right)
\quad\mbox{and}\quad
2F^2_{z\zb}=-g_{z\zb}\,\left(\frac{1}{R^2}-\phi^\+\phi \right)\ .
\end{eqnarray}
\end{subequations}
In the Abelian case $n=1$ and for $A^1=-A^2=\sfrac12\, A$ these equations reduce to the 
standard Abelian BPS vortex equations on $\Sigma$ (see e.g.~\cite{Taubes:1979ps, 
Bradlow, Popov:2007ms}) which can also be considered on $\Sigma =\R^2$.

\smallskip

\noindent
{\bf Topological restrictions.} Recall that $A^1$ and $A^2$ are connections on 
Hermitian vector bundles $E_1$ and $E_2$ over $\Sigma$ having rank $n$ and $1$,
respectively. Calculating their first Chern numbers
\begin{equation}\label{3.8}
N_1=\mbox{deg}E_1=c_1(E_1)=\frac{\im}{2\pi}\,\int_{\Sigma}\tr\, F^1
\quad\mbox{and}\quad
N_2=\mbox{deg}E_2=c_1(E_2)=\frac{\im}{2\pi}\,\int_{\Sigma} F^2
\end{equation}
and integrating traces of (3.7b) over $\Sigma$, we obtain (cf.~\cite{Bradlow})
\begin{equation}\label{3.9}
N_1 \le\frac{n}{\varkappa R^2 }\, (1-g)\quad\mbox{and}\quad
N_1+N_2=\frac{1}{\varkappa R^2 }\, (n-1)(1-g)\ .
\end{equation}
Here $g\ne 1$ and for $g=1$ (torus) one can derive similar (in)equalities.
Thus, there are the topological obstructions (\ref{3.9}) to the existence of
solutions to the vortex equations (\ref{3.7}).

\smallskip

\noindent
{\bf Non-Abelian Higgs models with $n_f=1$.} Here we want to compare the vortex equations 
(\ref{3.7}) with those considered in non-Abelian Higgs models on $\R^2$ with U$(n_c)$ gauge
fields and $n_f$ scalar fields in the fundamental representation of the group U$(n_c)$ (see
e.g.~\cite{group, group01, group1} and references therein). In the literature, $n_c$ is 
identified with the number of ``colours" and $n_f$ with the number of ``flavours". In the case 
of (\ref{3.7}), one has $n_c=n\ge 1$ and $n_f=1$. It is usually argued~\cite{group1} that it 
should be $n_f\ge n_c$ since there is no non-Abelian vacuum with vanishing gauge fields and 
constant Higgs fields if $n_f< n_c$.
However, this argument may be true only for $\Sigma =\R^2$. On the contrary, on compact Riemann
surfaces $\Sigma$ there are no constant sections $\phi$ (besides zero section) of the non-trivial
bundle $E=E_1\otimes\bar E_2$ and the vacuum has more complicated structure. Therefore, on Riemann 
surfaces the case $n_f=1$ is really non-Abelian and (\ref{3.7}) describes the simplest non-Abelian 
vortex equations.

For more easy comparing of (\ref{3.7}) with the non-Abelian vortex equations discussed in the 
literature~\cite{group1}, we rewrite them in the form 
\begin{subequations}\label{3.10}
\begin{eqnarray}
&\pa_{\zb}\phi + A^{su(n)}_{\zb}\phi + A^{u(1)}_{\zb}\phi =0 \ ,\\
&2g^{z\zb}F^{su(n)}_{z\zb}=\frac{1}{n}\,\tr (\phi\,\phi^\+)\cdot 1_n - \phi\,\phi^\+
\quad\mbox{and}\quad
g^{z\zb}F^{u(1)}_{z\zb}=\frac{1}{R^2} - \frac{n+1}{2n}\phi^\+\phi \ ,
\end{eqnarray}
\end{subequations}
where
\begin{subequations}\label{3.11}
\begin{eqnarray}
A^{su(n)}= A^1 - \frac{1}{n}\,(\tr A^1)\cdot 1_n
&\mbox{and}&
F^{su(n)}= F^1-\frac{1}{n}\, (\tr F^1)\cdot 1_n \ ,\\
A^{u(1)}=\frac{1}{n}\tr A^1 - A^2 &\mbox{and}&
F^{u(1)}= \diff A^{u(1)}\ .
\end{eqnarray}
\end{subequations}
One can also introduce the Abelian fields
\begin{equation}\label{3.12}
\tilde A^{u(1)}=\frac{1}{n}\tr A^1 + A^2 \quad\mbox{and}\quad
\tilde F^{u(1)}= \diff \tilde{A}^{u(1)}\ ,
\end{equation}
but for $n>1$ they are not independent since from (\ref{3.7}) it follows that
\begin{equation}\label{3.13}
g^{z\zb}\tilde F^{u(1)}_{z\zb}=\frac{n-1}{n+1}\Bigl (\frac{1}{R^2} - 
g^{z\zb} F^{u(1)}_{z\zb}\Bigr )= \frac{n-1}{2n}\,\phi^\+\phi \ .
\end{equation}

Recall that $A^{su(n)}+A^{u(1)}\cdot 1_n$ is a connection on the rank-$n$ complex vector 
bundle $E$ and $\phi\in H^0(\Sigma , E)$ is a section of this  bundle. Accordingly, the
topological restrictions (\ref{3.9}) can be rewritten for (\ref{3.10}) in the form
\begin{subequations}\label{3.14}
\begin{eqnarray}
0\ \le\ N&=& \frac{1}{n}\,N_1 - N_2\ =\ \frac{\im}{2\pi}\,\int_{\Sigma}\, F^{u(1)}\ \le\
\frac{2}{\varkappa R^2 }\, (1-g)\ ,\\
0\ \le\ \tilde N&=& \frac{1}{n}\,N_1+N_2\ =\ \frac{\im}{2\pi}\,\int_{\Sigma}\, \tilde F^{u(1)}
\ \le\ \frac{n-1}{n+1}\, \frac{2}{\varkappa R^2 }\,(1-g)\ ,
\end{eqnarray}
\end{subequations}
plus obvious equation connecting $N$ and $\tilde N$. For a given $g\ne 1$ (for tori one can 
derive similar formulae) all these restrictions can be satisfied 
by a proper choice of $\varkappa$.

Summarizing the above discussion, we have shown that for models with $n_f=1$ one can identify 
non-Abelian vortices on Riemann surfaces $\Sigma$ with SU(2)-equivariant instantons on manifolds
$\Sigma\times\C P^1$. It would be interesting to generalize these results to the case $n_f>1$ and 
arbitrary $n_c\ (=n)$.

\vspace{5mm}

\section{Integrability of non-Abelian BPS vortex equations}

\noindent
{\bf Integrable case.} We considered the BPS vortex equations (\ref{3.10}) on a Riemann surface $\Sigma$
and showed their equivalence to the instanton equations (\ref{3.6}) on the manifold $M=\Sigma\times\C P^1$. 
Note that one can rescale the metric on $\Sigma$,  $g_{z\zb}\to t^2g_{z\zb}$, so that
\begin{equation}\label{4.1}
R^t_{\Sigma}=\frac{2\varkappa}{t^2}=:2\varkappa_t
\end{equation}
and for $g>1$ one can always choose $t^2=-R^2/\varkappa$ (since $\varkappa <0$) so that
\begin{equation}\label{4.2}
R_M=R^t_{\Sigma}+R_{\C P^1}=0\ .
\end{equation}
In this case the Weyl tensor for the manifold $M$ is self-dual~\cite{B} and such manifolds are
considered as gravitational instantons in conformal gravity~\cite{Gibbons1, Berkovits}.

An important feature of K\"ahler manifolds with zero scalar curvature (\ref{4.2}) is that the so-called 
twistor space $\Zcal$ of $M$ becomes a complex manifold if (\ref{4.2}) is satisfied. Let us consider an 
open subset $\Ucal$ of $M=\Sigma\times\C P^1$ with complex coordinates $z, y$. Then the twistor 
space\footnote{For more details about twistor spaces see e.g.~\cite{Atiyah:1978wi}.} of $\Ucal$ (i.e. the 
restriction of $\Zcal$ to $\Ucal$) is diffeomorphic to $\Ucal\times\C P^1$,
\begin{equation}\label{4.3}
\Zcal|_{\Ucal}\simeq\Ucal\times\C P^1
\end{equation}
with a local complex coordinate $\la\in \C P^1\setminus\{\infty\}$ on the last factor. On $\Zcal$ there 
is a distribution generated by three vector fields of type (0,1) closed under the Lie bracket. They have 
the form~\cite{Popov:2007ms}
\begin{equation}\label{4.4}
V_{\1}:=\tilde e_{\1}-\la\tilde e_{2}\ ,\quad V_{\2}:=\tilde e_{\2}+\la\tilde e_{1}
\quad\mbox{and}\quad  V_{\3}=\pa_{\bar\la} \ ,
\end{equation}
where
\begin{subequations}\label{4.4'}
\begin{eqnarray}
\tilde e_1=\rho^{-1}\left(\pa_{z} - (\pa_{z}\log\rho)\la\pa_{\la}\right) \ ,\quad
\tilde e_{\1}= \rho^{-1}\left(\pa_{{\zb}} + (\pa_{\zb}\log\rho)\la\pa_{\la}\right)\ ,\\
\tilde e_2=\tilde \rho^{-1}\left(\pa_{y} - (\pa_{y}\log\tilde\rho)\la\pa_{\la}\right) \ ,\quad
\tilde e_{\2}= \tilde\rho^{-1}\left(\pa_{\yb} +(\pa_{\yb}\log\tilde\rho)\la\pa_{\la}\right) \ .
\end{eqnarray}
\end{subequations}
Recall that $\rho^2=g_{z\zb}$ and $\tilde\rho^2=g_{y\yb}$ are components of metrics on $\Sigma$ and 
$\C P^1$; the explicit form of $\tilde\rho$ is given in (\ref{2.4}). In the case (\ref{4.2}) one 
can pull back the instanton bundle $\Ecal\to M$ to the bundle $\hat\Ecal$ over the twistor space 
$\Zcal$ and introduce {\it integrable} holomorphic structure on $\hat\Ecal$ defined by a (0,1)-type 
connection along the vector fields (\ref{4.4}). The integrability of this structure is 
equivalent~\cite{Atiyah:1978wi, Ward:1977ta} to the self-duality equations (\ref{3.6}).

\smallskip

\noindent
{\bf Lax pair.} For the case (\ref{4.2}) from (3.14a) we obtain the inequality
\begin{equation}\label{4.5}
N \le 2\,(g-1)
\end{equation}
since $\varkappa_t=-1/R^2$. For having nonempty moduli space of solutions of
the vortex equations (\ref{3.10}) we assume that this condition is satisfied.

Let us now introduce an $n\times n$ matrix $\psi = \psi (x^{\m}, \la )$ which does not depend
on $\bar\la$ and consider two linear equations (Lax pair),
\begin{subequations}\label{4.6}
\begin{eqnarray}
\hat\nabla_{V_{\1}}\psi :=\bigl[\tilde{e}_{\1}+ \Acal_{\1}-{\la}\, (\tilde{e}_2 + 
\Acal_2)\bigr]\,\psi =0 \ ,\\
\hat\nabla_{V_{\2}}\psi :=\bigl[{\la}\, (\tilde{e}_{1}+ \Acal_{1}) +\tilde{e}_{\2} + 
\Acal_{\2}\bigr]\,\psi =0 \ ,
\end{eqnarray}
\end{subequations}
where $\tilde{e}_{1}, \tilde{e}_2, \tilde{e}_{\1}$ and $\tilde{e}_{\2}$ are written down in 
(\ref{4.4'}) and $\Acal_1, \Acal_2, \Acal_{\1}$ and $\Acal_{\2}$ can be easily extracted from
(\ref{3.1}). It is not difficult to check that the compatibility conditions of the linear
equations (\ref{4.6}), 
\begin{equation}\label{4.7}
[\hat\nabla_{V_{\1}},\hat\nabla_{V_{\2}}]\,\psi =0\ ,
\end{equation}
are equivalent to the vortex equations (\ref{3.10}). Thus, one can apply various 
solution-generating techniques (twistor approach, dressing method etc.) for solving 
non-Abelian BPS vortex equations on $\Sigma$ with the help of the Lax pair (\ref{4.6}).

\smallskip

To sum up, we have shown that the non-Abelian vortex equations on Riemann surfaces of 
genus $g>1$ are integrable under certain topological restrictions. It is of interest 
to generalize the constructions of this paper to supersymmetric vortex equations on 
$\Sigma$ since they will correspond to solutions of $\Ncal$-extended supersymmetric 
self-dual Yang-Mills theory~\cite{group2, group3} defined on the conformal gravitational 
instantons $\Sigma\times S^2$. Such vortices on $\Sigma$ will represent the simplest 
natural interaction of super-Yang-Mills instantons and gravitons in conformal (super)gravity 
appearing  in twistor string theory~\cite{Witten, Berkovits}. These Yang-Mills/gravity
configurations can be a good test background for calculation of open/closed twistor string 
amplitudes and the corresponding Yang-Mills/gravity amplitudes.

\end{document}